\documentclass{PoS}
\usepackage{subfigure}
\title{On the phenomenon of emergent spacetimes: \\
 An instruction guide for experimental cosmology}
\ShortTitle{Emergent spacetimes and experimental cosmology}
\author{\speaker{Silke Weinfurtner} \\
       Department of Physics and Astronomy \\ 
       University of British Columbia \\
       Vancouver, Canada \\
       E-mail: \email{silke@phas.ubc.ca}}
  \author{Matt Visser \\
       School of Mathematics, Statistics, and Computer Science \\ 
       Victoria University of Wellington \\ 
       New Zealand\\
       E-mail: \email{matt.visser@mcs.vuw.ac.nz}}
   \author{ Piyush Jain and C.~W. Gardiner. \\
Jack Dodd Centre for Quantum Technology \\
Physics Department \\
University of Otago \\
New Zealand \\
   E-mail: \email{piyushnz@gmail.com} $\,$ and $\,$ \email{gardiner@physics.otago.ac.nz}}
\newcommand{\refb}[1]{(\ref{#1})}
\abstract{
We present a toy model where spacetime is emergent from a more fundamental microscopic system, and investigate the gray area interpolating between the collective and free-particle regimes. For a period of rapid exponential growth in the analogue universe, we argue that the intermediate regime is best described by a coloured potpourri of geometries --- a ``rainbow geometry''. This can be viewed as an alternative approach towards understanding quantum field theories in the presence of Lorentz-symmetry breaking at ultraviolet scales. Firstly, it is pointed out that cosmological particle production in our emergent FRW-type analogue universe,  when compared to conventional semi-classical quantum gravity, is only temporarily robust against model-specific deviations from Lorentz invariance. Secondly, it is possible to carry out a straightforward quantitative analysis to estimate a suitable parameter regime for experimental (analogue) cosmology.
}
\FullConference{From Quantum to Emergent Gravity: Theory and Phenomenology\\
                June 11-15 2007\\
                Trieste, Italy}
\begin{document}
%%%%%%%%%%%%%
\def\d{{\mathrm{d}}}
%%%%%%%%%%%%%%%%%%%%%%%%%%%%%%%%%%%%%%%%%%%%%%%%%%%
%
\section{Introduction\label{Sec:Introduction}}
%
%%%%%%%%%%%%%%%%%%%%%%%%%%%%%%%%%%%%%%%%%%%%%%%%%%%
The idea of setting up laboratory-based toy models for quantum field theory has been discussed, at the very least, ever since Bill Unruh's development in 1981~\cite{Unruh:1981bi} of what are now known as ``dumb holes'' or ``acoustic black holes''. In that initial article,  Unruh first indicated the analogy between the motion of sound waves in a convergent fluid flow and massless spin-zero particles exposed to a black hole.

Gradually, the acoustic metric\,/\,analogue gravity\,/\,emergent spacetime programme has been extended to various media (e.g., for an electromagnetic waveguide~\cite{Schutzhold:2005aa}, Bose--Einstein condensates~\cite{Barcelo:2001gt}, and superfluid helium~\cite{Volovik:1995za}). Emergent spacetimes in superfluids are of special interest for experimental purposes. 
The extremely low background temperatures for superfluids enable us in principle to detect tiny quantum effects, such as Hawking radiation and ``externally driven'' particle production. In addition, the experimental techniques to control superfluids,  (for example Bose--Einstein condensates), are already at a very sophisticated level and further progress in quasi-particle detection mechanisms is expected~\cite{Fischer:2004iy,Schutzhold:2006pv,Weinfurtner:2007aa}. 
There are plenty of technical problems that have to be addressed before the experimental laboratory realization of black holes, or cosmological particle production, but the situation does not seem by any means to be hopeless.

In the following we will therefore focus on spacetimes emergent from an ultra-cold gas of Bosons. It has been shown, see for example~\cite{Barcelo:2003ia, Barcelo:2003yk, Fedichev:2004on, Fedichev:2004fi, Fischer:2004iy, Schutzhold:2005mc, Uhlmann:2005rx, Weinfurtner:2006eq}, that \textit{in principle} it is possible to manipulate the speed of sound through external fields~ to mimic the behavior of quantum modes in Friedmann--Robertson--Walker-type (FRW) universe. Previously, in~\cite{Jain:2006ki,Weinfurtner:2007ab}, we argued that before attacking specific problems involved with the experimental set-up, one needs to carefully choose a suitable parameter regime for such an experiment. In the following we would like to summarize and extend these ideas, and present an instruction guide for experimental analogue cosmology via Bose--Einstein condensates, see boxed text in Sec.~\ref{Sec:Instruction.Guid.for.Experimetal.Cosmology}. These ideas will be presented in the second half of our paper.

In the first half, we will comment on the phenomenology of emergent spacetimes, and address its relevance for quantum gravity and quantum gravity phenomenology. Systematically, we explain why the particle production process in parametrically excited condensates, corresponding to emergent Friedmann--Robertson--Walker geometries, is in general not robust against model-specific deviations from ``Lorentz invariance''. The modifications in the collective regime originate in the microscopic physics of the condensate, i.e., the fundamental Bosons. These modifications break the Lorentz symmetry in the analogue model at ultraviolet scales~\cite{Jacobson:1991sd}. A significant branch of so-called ``quantum gravity phenomenology''  focusses on the consequences of Lorentz violations at high energies, and thus the study of stability\,/\,robustness of semi-classical quantum gravity against ``Planck-scale modification'' is of great theoretical interest~\cite{Jacobson:1991sd,Unruh:1994zw,Unruh:1995aa}. We also suggest an alternative approach to get a grasp on the momentum-dependent behavior of quantum modes in an explicitly time-dependent external geometry, by using the notion of rainbow spacetimes.

%+++++++++++++++++++++++++++++++++++++++++++++++++++++++++++++++++
\subsection{Emergent spacetime geometries from ultracold Bose gases\label{Sec:Intro.Emergent.Spacetimes}}
%+++++++++++++++++++++++++++++++++++++++++++++++++++++++++++++++++
In the following we present a relatively simple and well-understood system, that is --- to some extent, as we will show in this paper --- capable of mimicking the behavior of quantum field modes exposed to an inflationary universe. 
The specific emergent spacetime geometry we are investigating exhibits the following key features:
\begin{description}
\item[High temperature phase:]{
At the most fundamental level we are dealing with non-hermitian quantum field operators representing the creation, $\hat\psi^\dag(t,\mathbf x)$, or destruction, $\hat\psi(t,\mathbf x)$, of an individual Boson at  a particular point in time and space. 
The relations between two field operators at $\mathbf x$ and $\mathbf x'$ are given by three equal time commutators
\begin{eqnarray}
\label{Eq:C1_1}&&\left[ \hat\psi(t,\mathbf x) , \hat \psi(t,\mathbf x') \right] = 0 , \\
\label{Eq:C2_1}&&\left[ \hat\psi^\dag(t,\mathbf x) , \hat \psi^\dag(t,\mathbf x') \right] = 0 , \\
\label{Eq:C3_1}&&\left[ \hat\psi^\dag(t,\mathbf x) , \hat \psi(t,\mathbf x') \right] = \delta(\mathbf x - \mathbf x') . 
\end{eqnarray}
For our purposes we consider a gas of trapped, ultra-cold, highly dilute and weakly interacting Bosons. Thus the Hamiltonian can be written as
\begin{equation}
\label{Eq:Fundamental.Hamiltonian}
\hat H = \int \d\mathbf x \left( -\hat \psi^\dag \frac{\hbar^2}{2m} \nabla^2 \hat\psi   
+ \hat \psi^\dag V_{\mathrm{ext}} \hat\psi 
+ \frac{U}{2} \hat\psi^\dag \hat\psi^\dag \hat\psi \hat \psi \right),
\end{equation}
the sum of the kinetic energy of the Boson field and the two potential energy contributions; the external trap, $V_{\mathrm{ext}}$, and the particle interactions, $U$. Due to the extreme dilution of the gas only two-particle interactions are taken into account, and in the weak-interaction regime the inter-atomic potential can be approximated by a pseudo-contact potential
\begin{equation}
U=\frac{4 \pi \hbar^2 a}{m}.
\end{equation}
Here $m$ is identified with the single-Boson mass, and $a$\, the $s$-wave scattering length. For this paper we only consider repulsive, $a>0$, inter-atomic forces. Experimentally, both negative and positive values for $a$ are accessible (by tuning external magnetic fields), and correspond to repulsive and attractive atomic interactions. It is interesting to notice that the nature of the microscopic interactions is related to the signature of the emergent spacetime. Repulsive (attractive) atom-atom interactions can be connected with a Lorentzian (Riemannian) spacetime signature, see for example~\cite{Calzetta1:2003xb,Calzetta:2005yk,Hu:2003,Weinfurtner:2007aa}.
 }
\item[Low temperature phase:]{
This is a regime where the microscopic degrees of freedom give way to macroscopic variables, such that the creation and destruction field operators can be replaced by classical mean fields, $\hat \psi^\dag \rightarrow \langle \hat\psi^\dag \rangle \equiv \psi^*$ and  $\hat \psi \rightarrow \langle \hat\psi \rangle \equiv \psi$.
For topologically trivial regions, without zeros or singularities, the complex macroscopic field may be written as
\begin{equation}
\psi(t,\mathbf x) = \sqrt{n(t,\mathbf x)} \; \exp(i\theta(t,\mathbf x)),
\end{equation} 
a function depending on two collective real-valued variables; the field amplitude as the square root of the condensate density, $n(t,\mathbf x)$, and an arbitrary (but fixed) phase, $\theta(t,\mathbf x)$.
Therefore, the single-particle Hamilitonian is no longer invariant under phase transformations of the kind $\theta \rightarrow \tilde\theta \exp(i\alpha)$. The $U(1)\equiv SO(2)$ symmetry of the Bose gas is broken spontaneously at the transition temperature $T_{c}$.
Below $T_{c}$ a large fraction of the atoms collapse into the lowest quantum state, and the gas undergoes a Bose--Einstein condensation. In this state of matter the quantum nature of the atoms becomes apparent on macroscopic scales, and the high and low temperature phases in our system are connected through a first-order phase transition associated with a spontaneous symmetry breaking.
}
\item[Semi-classical quantum geometry picture:]{In the hydrodynamic limit a geometrical rank two tensor can be identified, dominating the evolution of linearized classical and quantum excitations around the mean field, $\theta \rightarrow \theta_{0} + \hat \theta$ and $n \rightarrow n_{0} + \hat n$. The dynamical equations are
\begin{equation}
\label{Eq:KGE}
\frac{1}{\sqrt{\vert g \vert}} \; \partial_{a} \, \left(\sqrt{\vert g \vert} \, g^{ab} \, \partial_{b} \hat \theta \right)=0\, ,  
\end{equation}
and small density fluctuations are considered to be the conjugate momenta to small phase perturbations, $\hat n = \hat \Pi_{\hat\theta}=-\sqrt{\vert g \vert} \, g^{tb} \, \partial_{b} \hat \theta$, on the emergent\,/\,analogue\,/\,acoustic spacetime,
\begin{equation}
\label{Eq:Acoustic.Metric} 
g_{a b} \equiv \left(
\frac{n \, \hbar}{c \, m } \right)^{\frac{2}{d-1}} \,
\left[ \begin{array}{c|c} -(c^{2}-v^{2}) & -v^j \\ \hline -v^i & \delta^{ij} \end{array}\right] .
\end{equation}
The conformal factor depends on the spatial dimensionality $d$ of the condensate. The background velocity $\mathbf v$ is given by
\begin{equation} 
\label{Eq:Background.Velocity}
\mathbf v = \frac{\hbar}{m} \, \nabla \theta_{0} \, ,
\end{equation}
as the gradient of the condensate phase $\theta_{0}$, and $c$ denotes the speed of sound.
The transformations applied to the field operators preserve the initial commutation relations~(\ref{Eq:C1_1}-\ref{Eq:C3_1}):
\begin{eqnarray}
\label{Eq:C1_5}&& \left[ \hat \theta(t,\mathbf x),\hat\theta(t,\mathbf x')\right] = 0 \, , \\ 
\label{Eq:C2_5}&& \left[ \hat\Pi_{\hat\theta}(t,\mathbf x),\hat\Pi_{\hat\theta}(t,\mathbf x')\right] =0 \, ,\\ 
\label{Eq:C3_5}&& \left[ \hat \theta(t,\mathbf x),\hat\Pi_{\hat\theta}(t,\mathbf x')\right] = i \delta(\mathbf x - \mathbf x') . 
\end{eqnarray}

From a field theory point of view, the present model is only capable of mimicking spin-zero \emph{massless} scalar fields.
However, it is also possible to develop an analogy between multi-component Bose--Einstein condensates and \emph{massive} spin-zero scalar fields. The addition of extra fields is necessary, as the fundamental Hamiltonian~\refb{Eq:Fundamental.Hamiltonian} undergoes a spontaneous symmetry breaking at $T_{c}$, predicting at least one massless field excitation; see  e.g. the Nambu--Goldstone theorem in~\cite{Zee:2004aa}. In an $n$-component condensate we expect $n$ excitations, where $n-1$ of them can have a non-zero mass. A full treatment of a $2$-component Bose--Einstein condensate with respect to the analogue model programme can be found in~\cite{Weinfurtner:2006nl,Weinfurtner:2006eq,Weinfurtner:2006iv} and~\cite{Liberati:2006kw,Liberati:2006sj}.

Furthermore, there are also kinematical and dynamical differences between the emergent metric tensor~\refb{Eq:Acoustic.Metric}, and the gravitational metric tensor encountered in general relativity. Firstly, the emergent spacetime components are functions of the macroscopic mean field variables, and thus possess only two degrees of freedom. Therefore in comparison with general relativity --- where we are dealing with six degrees of freedom --- the analogy is only fully applicable in a limited number of (typically highly symmetric) spacetimes. 

The emergent spacetime picture is derived under the premise that field perturbations are negligibly small, $\langle \delta \hat\psi \rangle \equiv 0$ and $\langle \delta \hat\psi^\dag \rangle \equiv 0$, and therefore will not backreact with the classical mean field, $\langle \psi + \delta \hat\psi \rangle \approx \psi $ and $\langle \psi^* + \delta \hat\psi^* \rangle \approx \psi^*$. Within this approximation the system is \textit{in principle} capable of mimicking quantum field effects, where the gravitational field is retained as a merely classical background insensitive to the evolution of its quantum perturbations. Beyond the validity of this approximation the analogy seems to break down, as the dynamics of the emergent spacetime is governed by the second quantized Hamiltonian for a system of Bosons. In general the Hamiltonian~\refb{Eq:Fundamental.Hamiltonian} is only in some cases appropriate to describe the dynamics of the system, when three-body recombination effects can be neglected; a thorough treatment of higher-order terms involves a number-conserving approach as presented in~\cite{Morgan:1999aa,Gardiner:2006aa}. It can be shown that the dynamics of the emergent spacetime description, correlated to the ground state of the system, is different from Einstein's theory of gravity. This issue has been investigated in~\cite{Fischer:2005iy}.\footnote{We would like to point out that our specific model captures some --- but not all --- relevant physical ingredients necessary for \textit{quantum graphity}~\cite{Markopoulou:2007jf,Markopoulou:2007ha,Markopoulou:2007qg,Konopka:2006hu,Konopka:2008hp}. Generally, the analogue models\,/\,emergent spacetime programme might also be of phenomenological value for alternative approaches for quantum gravity involving some microstructure dominant at very small scales.}

As indicated in the previous paragraph, while the analogy is \textit{in principle} capable of mimicking quantum field theory effects in curved spacetime, any specific analogue model currently known will exhibit some specific corrections that \textit{might} lead to significant modifications to the particle production process. Surprisingly, the model specific alterations enter (to first order) in a relatively simple manner, and should be viewed as an essential part of the emergent spacetime picture.  
}
\end{description}
All of the above is derived under the premise of the \textit{hydrodynamic approximation}, where the so-called \textit{quantum pressure} in the superfluid is negligible, that is:
\begin{equation}
\label{Eq:Hydrodynamic.Limit}
\vert U \, n_{0} \vert  \gg \vert (\hbar^{2}/2m) \, \widetilde D_{2} \, n_{0} \vert  .
\end{equation}
This holds when the kinetic energy of density fluctuations in the condensate, $(\hbar^{2}/2m) \, \widetilde D_{2} \, n_{0} /  n_{0}$, is small compared to the atom-atom interaction strength, $U$. Here $\widetilde D_{2}$, a differential operator acting on $n_0$, is defined as
\begin{equation}
\label{QuantumPressure}
\widetilde D_{2} =\frac{1}{2} \left\{ \frac{(\nabla n_{0})^{2} -(\nabla^{2}n_{0})n_{0}}{n_{0}^{3}} -\frac{\nabla n_{0}}{n_{0}^{2}}\nabla +\frac{1}{n_{0}}\nabla^{2} \right\} .
\end{equation}
As the differential operator depends on spatial derivatives, it will increasingly alter the behavior of quantum field modes with higher wavenumbers, $k$. Luckily, all modifications are formally taken into account  by simply replacing the atom-atom interactions $U$ by a differential operator $\widetilde U$:
\begin{equation}
\label{Eq.Effective.U}
\widetilde U = U - \frac{\hbar^2}{2 m}\;\widetilde D_{2}   \, .
\end{equation}
This leads to modified hydrodynamic equations involving non-trivial implications for the emergent spacetime programme. 
\begin{description}
\item[Non-perturbative ultra-violet corrections:]{
Now set  $n_0(t,\mathbf x)\equiv n_0$ and $ \theta_0(t,\mathbf x) \equiv \theta_0$, so that the emergent metric~\refb{Eq:Acoustic.Metric} is equivalent to that of Minkowski spacetime. Then the dispersion relation for small classical and quantum fluctuations around this classical ground state  is given by 
\begin{equation}
\label{Eq:Omega.NonLinear.Minkowski}
\omega_{k}^2=
c^{2}k^{2} + \gamma_{\mathrm{qp}}^{2}k^{4} \equiv c^2 k^2  \left( 1 + \frac{k^2}{K^2}  \right),
\end{equation}
where it is useful to define the quantities
\begin{equation}
\label{qp}
\gamma_{\mathrm{qp}}=\frac{\hbar}{2m} , \quad \mbox{and} \quad K= \frac{c}{\gamma_\mathrm{qp}},
\end{equation}
such that we easily obtain the hydrodynamic limit, for $\gamma_{\mathrm{qp}} \rightarrow 0$, or alternatively $K \rightarrow \infty$. (For a detailed derivation see for example~\cite{Weinfurtner:2007ab,Jain:2006ki}).

Thus, the present system exhibits an emergent Lorentz symmetry for infrared quantum modes,
\begin{equation}
k_{\mathrm{IR}} \quad  : \quad \vert k \vert \ll K ,
\end{equation}
where effectively $\gamma_{\mathrm{qp}} \rightarrow 0$, and the hydrodynamic limit is applicable. 
At small scales --- relative to $1/K$ --- this symmetry is broken, as model specific corrections become apparent for crossover and ultraviolet modes,
\begin{equation}
k_{\mathrm{crossover}} \quad  : \quad \vert k \vert \sim K, \qquad\qquad k_{\mathrm{UV}} \quad  : \quad \vert k \vert \gg K, 
\end{equation}
consequently we must keep $\gamma_{\mathrm{qp}} \neq 0$ in these regimes, and quantum pressure effects of the superfluid significantly influence the behavior of small fluctuations.
In this spirit it is plausible to introduce an \textit{analogue Planck-length},
\begin{equation}
\ell_{\mathrm{Planck}} = \frac{\gamma_\mathrm{qp}}{c} \equiv \frac{\hbar}{2m \, c} \equiv \frac{1}{K}.
\end{equation}
The reader may consider that in some sense the \textit{emergent Lorentz invariance breaking} (LIV) \textit{scale}  has to be correlated with ``new physics'' and we will further advocate this point of view below.

Summarizing the above, we see that emergent spacetimes\,/\,analogue models exhibit an emergent\,/\,effective Lorentz symmetry for low-energy\,/\,infrared excitations around the macroscopic mean field. This symmetry will be broken in the high-energy\,/\,ultraviolet regime, that is at scales where collective classical and quantum fluctuations first experience effects from the underlying microscopic theory. The present model focuses on the boost subgroup that supports CPT invariance and results in a momentum-dependent dispersion relation. These corrections originate in the hydrodynamic fluid equations, and hence are of non-perturbative nature.}
\end{description}
We trust that the above has provided readers unfamiliar with the analogue models programme with the key features necessary to understand the parallelism. Analogue models are a generic tool for probing the interface between gravity and quantum physics. From now on we would like to restrict our toy model further, and focus on the possibility of mimicking cosmological spacetimes.

%+++++++++++++++++++++++++++++++++++++++++++++++++++++++++++++++++
\subsection{FRW-type spacetime geometries and degrees of freedom\label{Sec:Mimicking.FRW.Spacetimes}}
%+++++++++++++++++++++++++++++++++++++++++++++++++++++++++++++++++
In the hydrodynamic limit the speed of sound in a condensate with an explicit time-dependence, but still ``at rest'' --- zero background velocity,  $\mathbf v = 0$ --- can be expressed by
\begin{equation} 
c(t,\mathbf{x})^2 = \frac{4\pi \hbar^2}{m^2}  n(t,\mathbf{x})  a(t,\mathbf{x})\to  c_0(\mathbf{x})^2  b_{n}(t) \, b_{a}(t).
\end{equation}
Both the scattering length $a(t)=a_0  b_{a}(t)$ and the condensate density $n(t,\mathbf x)=n_0(\mathbf{x}) b_{n}(t)$ are allowed to vary with respect to laboratory time $t$. The initial condensate parameters, at the beginning of the experiment $t=t_0$, are given by $a_0$ and $n_0$, such that $b_a(t_0)=1$ and $b_n(t_0)=1$. Without any loss of generality we can set $t_0=0$. (Notice, that for the cases where $n_0$ exhibits a spatial dependence, one has to give up on a uniform sound-cone structure  throughout the condensate. In these scenarios the notion of FRW spacetime has to be restricted to an area where $n_0(\mathbf{x}) \approx n_0$ is approximately constant.)
Implementing this parameterization into the line-element (based on the metric given in equation~\refb{Eq:Acoustic.Metric}) we have
\begin{equation}  
\label{Equ:Line.Element.FRW.1}
    \d s^2 = \left(\frac{n_{0}}{c_{0}}\right)^{\frac{2}{d-1}} 
    \left[ - c_0^2 \, b_{n}(t)^{\frac{d}{d-1}}  b_{a}(t)^{\frac{d-2}{d-1}} \d t^2 
    +    \left(\frac{b_{n}(t)}{b_a(t)}\right)^{\frac{1}{d-1}} \d\mathbf{x}^2 \right] .
\end{equation}
Let us implement a change of coordinates $d\tau^2 = \, b_{n}(t)^{\frac{d}{d-1}}  b_{a}(t)^{\frac{d-2}{d-1}} dt^2$, such that
\begin{equation} \label{Equ:Line.Element.FRW.2}
\d s^2 = \left(\frac{n_{0}}{c_{0}}\right)^{\frac{2}{d-1}} \left[ - c_0^2 \, \d\tau^2 +   \left(\frac{b_{n}(\tau)}{b_a(\tau)}\right)^{\frac{1}{d-1}} \d\mathbf{x}^2 \right] ,
\end{equation}
where it is obvious that effectively --- in this parameterization for zero background velocity --- we are left with one degree of freedom, $g(\tau)= b_{n}(\tau)/b_a(\tau)$. By inspection this metric represents a spatially flat ($k=0$) FRW cosmological spacetime with scale factor
\begin{equation}
a_{\scriptscriptstyle\mathrm{FRW}}(t) = a_{\scriptscriptstyle\mathrm{FRW},0} \; 
\left(\frac{b_{n}(\tau)}{b_a(\tau)}\right)^{\frac{1}{2(d-1)}}.
\end{equation}

However, in the specific analogue spacetime under current investigation the situation is a more elaborate one as the applicability of this interpretation hinges on the validity of the hydrodynamic limit, or in the language of effective field theories on the (in this particular case time-dependent) effective Planck-length, given (in units of laboratory distance) by
\begin{equation}
\ell_{\mathrm{Planck}}(t) = \frac{\gamma_\mathrm{qp}}{c(t)} = 
\frac{\gamma_\mathrm{qp}}{c_0 \, \sqrt{b_{n}(t) b_a(t)}} = 
 \frac{ \ell_{\mathrm{Planck},0} }{\sqrt{b_{n}(t) b_a(t)}}.
\end{equation}
In addition, up to the present time we are lacking a thorough treatment of the possible modifications to the particle production process arising from, strictly speaking,  non-linear dispersion relations for spin-zero massless scalar fields in time-dependent parametrically excited analogue models. (This is in contrast to several analyses of dumb hole evaporation~\cite{Jacobson:1991sd, Jacobson:1993ab, Unruh:1994zw, Unruh:1995aa, Unruh:2005aa} focusing on this issue.) This raises the question of the ``robustness'' of particle production in effective spacetimes with time-dependent preferred-frame effects.

%+++++++++++++++++++++++++++++++++++++++++++++++++++++++++++++++++
\subsection{On the existence of geometry beyond the hydrodynamic limit\label{Sec:Existence.Geometry}}
%+++++++++++++++++++++++++++++++++++++++++++++++++++++++++++++++++
Consider a classical\,/\,quantum mode with wavelength $k$, assuming that at a particular time $t_1$ it is insensitive to quantum pressure effects in the condensate, $k \ll 2\pi / \ell_\mathrm{Planck} (t_1)$. Let us further assume that as the system evolves, we can find a time $t_2$, with $t_2 > t_1$, such that $k \gg 2\pi / \ell_\mathrm{Planck} (t_2)$, concluding that at $t_2$ the analogy has \textit{broken down}. There seems to be no caveat here, as long as one tries to explore the intermediate regime, where $t_1 < t '< t_2$, such that the wavenumber of the modes gradually changes from $k \ll 2\pi/\ell_\mathrm{Planck}(t_1)$ to $k \sim 2\pi/\ell_\mathrm{Planck}(t')$, and finally to $k \gg 2\pi/\ell_\mathrm{Planck}(t_2)$. There is in this situation no such thing as Einstein dynamics for the condensate parameters, nor is it possible to uniquely separate the notion of spacetime from the field equations, as in our model both arise simultaneously. This leaves us with the possibility of treating modifications resulting in non-linear terms in the dispersion relation as part of the geometry.

Let us briefly map out the technical steps involved in obtaining a geometrical interpretation for the kinematical behavior for collective perturbations in the condensate beyond the hydrodynamic limit.
For that we require the existence of:
\begin{enumerate}
\item[\textbf{(i)}] hydrodynamic fluid equations, 
\item[\textbf{(ii)}] the integral differential operator $\tilde{U}^{-1}$, and finally 
\item[\textbf{(iii)}] a relation between $f^{ab}$ \textit{and} $g^{ab}$.
\end{enumerate}
Regarding \textbf{(i)}, as mentioned previously quantum pressure effects are easily taken into account by formally replacing the atom-atom interaction variable with an interaction differential operator $U \rightarrow \tilde{U}$. In this spirit the \textit{modified} hydrodynamic fluid equations for classical\,/\,quantum perturbations are given by,
\begin{eqnarray} 
\label{Eq:on_dynamics}
\mbox{(Continuity equation)}&& \partial_t \hat n+\nabla \cdot \left[\left(\frac{n_{0} \hbar}{m}\nabla\hat\theta\right)+(n_{0} \, \mathbf v)
\right] =0\, ,\\
\label{Eq:otheta_dynamics}
\mbox{(Euler equation)}&&\partial_t \hat \theta+\mathbf v \cdot \nabla\hat\theta+\frac{\widetilde U}{\hbar}\;\hat n =0 \, .
\end{eqnarray}
Notice the modifications only enter the Euler equation, while the continuity equation remains untouched. The nomenclature ``quantum pressure'' originates from the fact that this modification is adding terms involving gradients of $\hat n$ to the fluid equation.\\
Regarding 
\textbf{(ii)}, to extract the analogy between fluid mechanics and classical\,/\,quantum field theory beyond the hydrodynamic limit it is necessary to find an explicit and tractable expression for the differential operator $\tilde{U}$. Only then are we able to merge the Euler and continuity equations into the form $\partial_a ( f^{ab} \partial_b \hat\theta ) = 0$. Since $\widetilde D_{2}$ and $\widetilde U$ are second-order linear differential operators, the inverse $\widetilde U^{-1}$ always exists as an integral operator (that is, in the sense of being a Green function). \\
Finally, \textbf{(iii)}, the $d+1$ dimensional matrix $f^{ab}$ derived from the modified hydrodynamic equations is contains inverse-differential-operator-valued entries:
\begin{equation}
\label{Eq:f}
f^{ab} = \hbar
\left[ \begin{array}{c|c} \vphantom{\Big|} -\widetilde U^{-1} & -\widetilde U^{-1} v^{j} \\ \hline \vphantom{\Big|} -v^{i} \widetilde U^{-1} & \frac{n_{0}}{m}\delta^{ij}- v^i \widetilde U^{-1} v^j \end{array}\right] .
\end{equation}
Notice that in general $ -\widetilde U^{-1} v^{i} \neq -v^{i} \widetilde U^{-1}$, and thus quantum pressure effects may at first glance seem to introduce non-symmetric effects in the spacetime geometry. However these two objects are Hermitian adjoints in the sense of integral operators, and so $f^{ab}$ is formally self--adjoint when acting on the appropriate function space, which (rather than naive symmetry) is the key property that one really wishes to preserve.
 In addition, we need to find an (inverse) metric tensor $g_{ab}$ such that $f^{ab} \equiv \sqrt{-g} \, g^{ab}$ where $g$ is the determinant of $g_{ab}$. Only then is the connection formally made to the field equation for a minimally
coupled massless scalar field in a curved spacetime. However, this last step is by no means obvious, as $\tilde{U}^{-1}$ does not necessarily commute with the $v^j$'s and\,/\,or $n_0$.

It goes without saying that only in the hydrodynamic limit, when $\tilde{U}^{-1}$ is replaceable with $U$, do we fully recover a  conventional spacetime geometry in our condensed matter system. Going beyond the hydrodynamic limit, to some extent the so-called \textit{eikonal limit} is applicable, leading to the notion of \textit{rainbow spacetimes}.
%

%~~~~~~~~~~~~~~~~~~~~~~~~~~~~~~~~~~~~~~~~~~~~~~~~~~~~
\subsubsection{The notion of rainbow spacetimes\label{Sec:Notion.Rainbow.Spacetimes}}
%~~~~~~~~~~~~~~~~~~~~~~~~~~~~~~~~~~~~~~~~~~~~~~~~~~~~
In the \emph{eikonal limit}, the differential operator  $\widetilde U$ can usefully be approximated by a function $\widetilde U \to U_{k}(t,\mathbf x) = U(t,\mathbf x) + \frac{\hbar^2 k^2}{ 4 m n_0} $, which we shall conveniently abbreviate by writing $U_k$.
Beyond the hydrodynamic limit, but within the eikonal approximation,  we obtain
\begin{equation}
\label{f2}
f^{ab} = \frac{\hbar}{U_k} 
\left[ \begin{array}{c|c} \vphantom{\Big|} -1 & - v^{j} \\ \hline \vphantom{\Big|} -v^{i}  & \frac{n_{0} U_k}{m}\delta^{ij}- v^i  v^j \end{array}\right]
\, .
\end{equation} 
Note that in the eikonal approximation the $k$ dependence hiding in $U_k$ will make this a momentum-dependent quantity, leading to a so-called \emph{rainbow metric}. It is convenient to define a momentum dependent speed of sound $c_{k}(t)^{2}= n_0 U_k/ m$ and so write
\begin{equation}
\label{f2b}
f^{ab} = \left( \frac{n_0 \hbar}{c_k^2 m}\right) 
\left[ \begin{array}{c|c} \vphantom{\Big|} -1 & - v^{j} \\ \hline \vphantom{\Big|} -v^{i}  & c_k^2 \delta^{ij}- v^i  v^j \end{array}\right]
\, .
\end{equation} 
The metric tensor is explicitly given by
\begin{equation}
\label{Eq:Rainbow.Metric}
g_{a b} \equiv
\left( \frac{n_{0}\, \hbar}{c_k \, m} \right)^{\frac{2}{d-1}} \,
\left[ \begin{array}{c|c} -(c_k^{2}-v^{2}) & -v^j \\ \hline -v^i & \delta^{ij} \end{array}\right] \, ,
\end{equation}
where we observe that  $c_{k}(t)^{2}=c(t)^{2}+\gamma_{\mathrm{qp}}^{2} k^{2}$.
Before we move on and exemplify the usefulness of the notion of rainbow spacetimes, we would like to point out that in the special case where the all entries in $f^{ab}$ only depend on the laboratory time $t$, but not on spatial coordinates $\mathbf{x}$, then symmetry and the existence of $g^{ab}$ is guaranteed.

The introduction of rainbow spacetimes might seem at first glance merely to be an artificial and unnecessary way to complicate the situation with the analogue models. The standard route to investigate the behavior of Lorentz symmetry breaking effects at ultraviolet scales is to study effective field theories. However, sometimes we are not able to solve the full problem, and that is when rainbow spacetimes can play an important role --- as they give us an alternative insight towards the understanding of particle production in a FRW-like spacetime under the presence of Planck-scale modifications. 
%%%%%%%%%%%%%%%%%%%%%%%%%%%%%%%%%%%%%%%%%%%%%%%%%%%
%
\section{Experimental cosmology by way of illustration\label{Sec:Cosmology.Tunable.BECs}}
%
%%%%%%%%%%%%%%%%%%%%%%%%%%%%%%%%%%%%%%%%%%%%%%%%%%%

In the following we will focus on a very simple, and in some sense idealized realization of laboratory cosmology. Here the novelty lies in the possibility of comparing our model presented above with explicit numerical simulations~\cite{Jain:2006ki,Weinfurtner:2007ab}; simulations that are purely based on the particular condensed matter system, without imposing any of the assumptions necessary to derive the emergent spacetime picture. 

The analogy we are interested in is quantum field theory in spatially flat $k=0$ Friedmann--Robertson--Walker geometries in two spatial dimensions $d=2$,
\begin{equation}
\label{Eq:FRW.GR}
\d s^{2} = g_{ab} \; \d x^{a} \, \d x^{b}= -\d\tau^{2} + a(\tau)^{2} \sum_{i=1}^{2} (\d x^{i})^{2} . 
\end{equation}
This is technically imposed by assuming a uniform and constant condensate density, $n(t,\mathbf{x})\equiv n_0$, but allowing the scattering length to vary in time. Altogether we set $b_n(t)=1$ and $b_a(t)= b(t)$, such that the effective line-element~\refb{Equ:Line.Element.FRW.1} reduces to
\begin{equation}
\label{met2Dconstrho}
\mathrm{d} s^2 
= \left( \frac{n_0}{c_0}\right)^{2} \left[ -c_0^2 \,  \d t^2+ b_{k}(t)^{-1} \d\mathbf x^2 \right] .
\end{equation}
For two spatial dimensions laboratory time $t$ and proper time $\tau$ are of the same form, and for the momentum-dependent FRW scale factor we obtain the very simple relationship,\footnote{A more detailed treatment can be found in \cite{Jain:2006ki,Weinfurtner:2007ab} and in \cite{Barcelo:2003ia,Barcelo:2003yk}. Alternatively in~\cite{Fedichev:2004on,Fedichev:2004fi,Fischer:2004iy,Uhlmann:2005rx,Unruh:1981bi} the authors kept the atom-atom interactions constant $b_a(t)=1$, working instead with time- and space-dependent condensate density. One then encounters non-uniformal sound-cone structure and (especially for the cases of a freely expanding condensate) a destructive measurement set-up where back-reaction effects might turn out to be significant.}
\begin{equation}
\label{Eq:FRWsaclefatcor_2d}
a_{k}(t) = b_{k}(t)^{-1/2}=\frac{1}{\sqrt{b(t) + (k/K)^{2}}} .
\end{equation}
%

%+++++++++++++++++++++++++++++++++++++++++++++++++++++++++++++++++
\subsection{Emergent rainbow inflation\label{Sec:Emergent.Rainbow.Inflation}}
%+++++++++++++++++++++++++++++++++++++++++++++++++++++++++++++++++
%
%_figure__figure__figure__figure__figure__figure__figure__figure__figure__figure__figure__figure_
\begin{figure*}[!htb]
\begin{center}
\mbox{
\subfigure[$\,$ Scale factor including quantum pressure effects;  \break \null \qquad with timescale  $t_{s}=1 \times 10^{-5}$. 
\label{Fig:FIG_desitter_ts1e-05_4000000_200000_SF_-55_25}]{\includegraphics[width=2.8in]{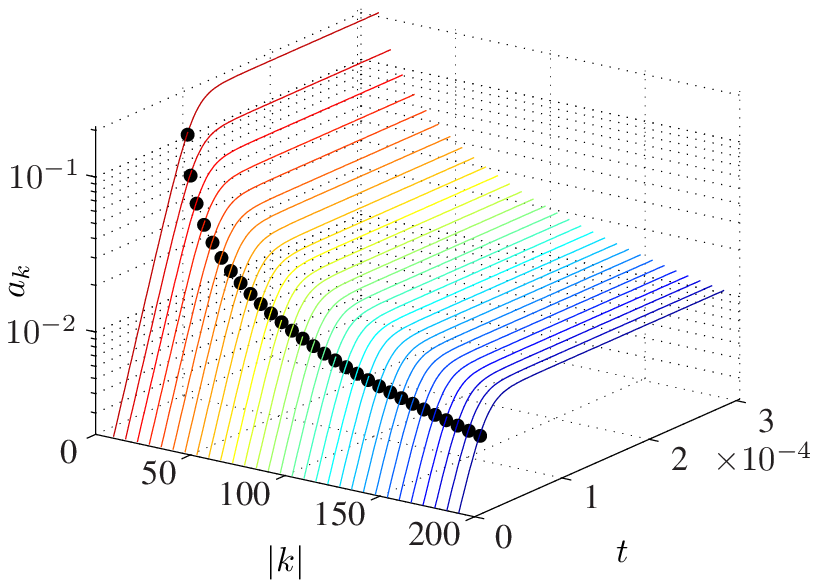}}
\hspace{0mm}
\subfigure[$\,$ Scale factor including quantum pressure effects; \break  \null \qquad with timescale $t_{s}=1 \times 10^{-5}$. 
\label{Fig:FIG_desitter_ts1e-05_4000000_200000_SF_0_0}]{\includegraphics[width=2.8in]{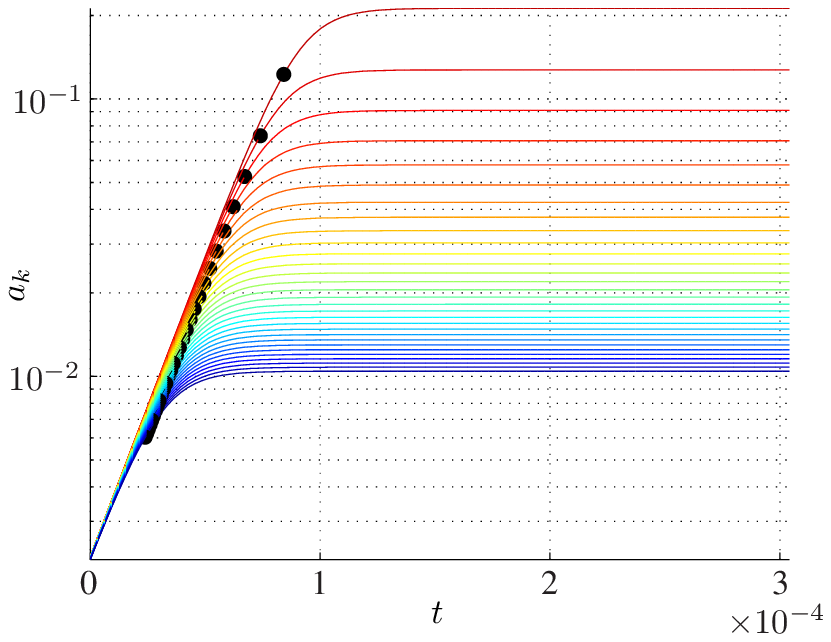}}     
}
\caption[Rainbow scale factor for effective inflation in Bose gas.]{(Colors online only.) In this figure we plot the logarithm of the scale function $a_{k}(t)$ for each $k$-value --- for $k \in [9,191]$ --- in a different colour. The different colours encode the energy of the modes: Gradually changing from low-energy\,/\,infrared (dark red) to high-energy\,/\,ultraviolet (dark blue). 
While the rainbow-scale factor approaches that of the hydrodynamic limit for low-energy modes, the ultraviolet modes show strong deviations. Note, that in the infinite past all modes are phononic, and therefore $a_{k}(t)\to a(t)$. The black dots indicate the time-dependent crossover (phononic to trans-phononic) in every quantum mode. Parameters are $C_{NL}(\bar{t}=0)=2 \times 10^{5}$, $N_{0}=10^{7}$ and $X=4 \times 10^{6}$.  (See~\cite{Jain:2006ki,Weinfurtner:2007ab} for details of the simulations.)}
\label{Fig:deSitter.Rainbow.ScaleFactor}
\end{center}
\end{figure*}
%_figure__figure__figure__figure__figure__figure__figure__figure__figure__figure__figure__figure_
%

Perhaps the most interesting cosmological case to study in our emergent spacetime is the \emph{de Sitter} universe, where the scale factor is given by an exponentially expanding (or contracting) universe, $a(\tau)=\exp(H \tau)$.
The concept of cosmological inflation was introduced simultaneously around 1981 and 1982 by Guth~\cite{Guth:1981aa}, Linde~\cite{Linde:1990aa}, and Albrecht and Steinhardt~\cite{Albrecht:1982aa} to explain the homogeneity of the temperature observed in our universe, beyond casually disconnected areas. Not long after (see, e.g., Guth~\cite{Guth:1981aa}, Hawking~\cite{Hawking:1982aa}, Bardeen~\cite{PhysRevD.22.1882}, Turner~\cite{Turner:1993aa} and Brandenburger~\cite{Brandenberger:1983aa}) it was realized that inflation also accounts for the existence of the perturbations in our universe today.

By now it should be clear that in order to simulate the behavior for quantum modes exposed to an inflationary universe in our $2$-dimensional superfluid, we shall have to make some compromises. In particular, we choose the scale factor for the atomic interactions to be $b(t)=\exp(-t/t_s)$, such that in the hydrodynamic limit the model is approaching the de Sitter case, $a_k(t) \rightarrow a(t)$. Unfortunately, that is the best one can do, one cannot make all momentum modes simultaneously see the same de Sitter universe.

The specific scale factor in the emergent line-element (including quantum pressure effects) is 
\begin{equation}
\label{Eq:Rainbow.scale.factor.deSitter.eikonal}
a_{k}(t) =\frac{1}{ \sqrt{\exp(-2H \, t)+ (k/K)^{2}} }.
\end{equation}
Thus the hydrodynamic, $\exp(-2H \, t) \gg \vert k/K \vert^{2}$, crossover,
$\exp(-2H \, t) \sim  \vert k/K \vert^{2}$, and free particle,
$\exp(-2H \, t) \ll  \vert k/K \vert^{2}$, limits are a matter of dividing the spectrum into appropriate energy regimes \emph{at a particular time} $t$. 
It is interesting that for early times --- when the interactions between the atoms are strong --- we naturally approach the hydrodynamic case, $\lim_{t \to - \infty} a_{k}(t) \to a(t)$, in the sense that most modes are phononic, and therefore larger and larger $k$-ranges are covered by ``conventional'' FRW-type quantum-field-theory. 

Quite the contrary occurs after an infinitely long-lasting expansion, where all modes behave as free particles, $\lim_{t \to + \infty} a_{k}(t) \to \vert K/k \vert \,$, and the universe, as seen by a mode with the wavelength $k$, will effectively approach a final finite fixed (momentum-dependent) size.

Before we continue, we wish to illustrate --- for the particular parameters used in our numerical simulations for a de Sitter-like universe --- where the phrase rainbow spacetime comes from; see Fig.~\ref{Fig:deSitter.Rainbow.ScaleFactor}. There we plot the emergent rainbow scale factor $a_{k}(t)$ for each $k$ mode using different colours --- gradually changing from dark red for infrared modes to dark blue for ultraviolet modes. The resulting colour-spectrum is reminiscent of on the colour spectrum obtained from real rainbows.

Due to this fundamental difference between our analogue model and the ``theory'' we wish to mimic, we know already that there will only be a finite time-period --- its length depends on the existence of the phononic regime, and therefore on the tunable initial interaction strength $U(0)=U_{0}$ --- beyond which the analogy breaks down. Note that the particle production process naturally ceases when the expansion rate slows down to zero. Thus we are facing a significant mathematical  problem, one that cannot be treated in a fully analytical manner as we pointed out previously in~\cite{Jain:2006ki,Weinfurtner:2007ab}. Fortunately, the  emergent rainbow spacetime picture is sufficient to provide good estimates for the particle production.

%+++++++++++++++++++++++++++++++++++++++++++++++++++++++++++++++++
\subsection{Instruction guide for experimental cosmology\label{Sec:Instruction.Guid.for.Experimetal.Cosmology}}
%+++++++++++++++++++++++++++++++++++++++++++++++++++++++++++++++++
The process of cosmological particle production in an expanding\,/\,collapsing universe can be qualitatively understood in terms of a single parameter, sepcifically, the frequency ratio $\mathcal{R}_{k}(t)$, see box. 
Initially, as pointed out in \cite{Weinfurtner:2007ab}, we derived this connection between the qualitative behavior of the particle production process and this frequency ratio in the hydrodynamic limit, where exact analytical studies are possible. We then extended  these ideas to the emergent rainbow metrics.

\begin{center}
\noindent
\fbox{
\begin{minipage}[t]{0.95\linewidth} 
We propose a simple $3$-step process to get a quantitative estimate of the particle production in our emergent spacetime, as a road map for experimental cosmology: 
\begin{itemize}
\item[\textbf{1} --]{Assign the initial condensate parameter, specifically the speed of sound $c_0=c_0(U_0,n_0,m)$, and the scaling functions for the scattering length $b(t/t_s)$. For a de Sitter-type universe the scaling time $t_s$ and the Hubble frequency $H$ are related by $H=1/(2 \, t_s)$.}
\item[\textbf{2} --]{Using the modified scale factor $a_k(t)$, see equation~\refb{Eq:Rainbow.scale.factor.deSitter.eikonal}, and the modified dispersion relation, here $\omega_k(t)=\omega_0 \sqrt{\exp(-2Ht)+(k/K)^2}$, compare with equation~\refb{Eq:Omega.NonLinear.Minkowski},  we compute the rainbow Hubble parameter $H_k(t)$:
\begin{equation}
H_k(t) := \frac{\dot a_k(t)}{a_k(t)} = H \; \frac{\exp(-2H \, t)}{\exp(-2H \, t)+ (k/K)^{2}} .
\end{equation}
Identify the ratio between the modified dispersion relation and the effective Hubble parameter as the significant parameter determining the particle production in our emergent spacetime,
\begin{equation}
\label{Eq:Ratio.deSitter.eikonal}
\mathcal R_{k}(t) =\frac{\omega_{k}(t)}{H_{k}(t)}= \frac{\omega_{0}}{H}  \; \frac{(\exp(-2H \, t) +  (k/K)^{2} )^{3/2} }{\exp(-2H \, t)} .
\end{equation}
Note that within the hydrodynamic limit $\mathcal R_{k}(t)  \rightarrow R_k(t) =  \frac{\omega_{0}}{H}  \; \exp(-H \, t) $, and $H_k(t) \rightarrow H$. As a rough rule of thumb we summarize: The smaller the frequency ratio the higher the final occupation number of the corresponding quantum mode.
}
\item[\textbf{3} --]{Estimate particle production by analyzing  $\mathcal R_{k}(t)$. A quantum mode with the wavenumber $k$ only experiences a significant amplifications when $\mathcal R_{k}(t) \ll 1$.}
\end{itemize}
\end{minipage} } 
\end{center}

It is a well-known result that the solutions for a spin-zero massless scalar field exposed to a de Sitter-type universe are a linear combination of first order Hankel functions of the first and second kind~\cite{Birrell:1984aa,Fulling:1989aa}. These mode functions are a function of the dimensionless ratio $R_{k}(t)=\omega / H$, and in the limit of $R_{k}\to \infty$ the mode functions approach ``freely oscillating'' positive and negative frequency modes, while for $R_{k} \to 0$ the modes stop oscillating, and the modes exhibit exponentially growing or exponentially decaying kinematics. The situation is lightly more complicated in our specific rainbow spacetime, that we wish to use to mimic cosmological inflation. However, we are able to define a modified frequency ratio, see box, which remains the only significant parameter in the system. 

\emph{Ex ante} we would like to motivate this section by the remark that while the frequency ratio in the hydrodynamic limit $R_{k}(t)$ is a monotonically decreasing function in time, the ratio in the eikonal approximation $\mathcal{R}_{k}(t)$ is not. 
Therefore there is some freedom to obtain results different from the ``conventional'' particle production process. 
We demonstrate the correctness of this assertion by referring to the numerical simulations reported in \cite{Jain:2006ki}.
\\

To obtain a rough estimate on the different qualitative regimes of particle production, we use the experience gained in the hydrodynamic limit, and simply exchange $R_{k}(t) \to \mathcal{R}_{k}(t)$. 

For early times, when $R_{k}(t)\gg 1$ the hydrodynamic and eikonal ratios are identical, and therefore in both cases approach the adiabatic limit, where we expect the particle production process to be negligibly small. 
Here quantum modes are approximately plane waves, but their amplitude and frequency change as a function of time.  This ansatz is referred to as the WKB approximation, which is valid within the \emph{adiabatic} limit,
when during one oscillation period $T=2\pi/ \omega_{k}(t)$ the relative change in the frequency is small (see \cite{Mukhanov:2007aa}),
$ \left\vert \frac{\omega_{k}(t+T)-\omega_{k}(t)}{\omega_{k}(t)} \right\vert \approx  
2\pi \, \left\vert \frac{\dot \omega_{k}}{\omega_{k}^{2}} \right\vert \ll 1 $. For de Sitter spacetimes equates to
$ \left\vert \frac{\dot \omega_{k}^{\mathrm{dS}}}{(\omega_{k}^{\mathrm{dS}})^{2}} \right\vert 
=   \left\vert \frac{1}{R_{k}(t)}  \right\vert  \ll 1 $,
the condition that the ratio $R_{k}(t)$ is much larger than one, which verifies the consistency of adopting the adiabatic approximation.

As intimated, the overall slope of the eikonal ratio is not a monotonically decreasing function, since
$\dot{\mathcal{R}}_{k}$ changes its sign at  $t_{\mathrm{turn}} = \frac{\ln (K^{2}/(2\, k^{2}))}{2H}$ .
For $t < t_{\mathrm{turn}}$ the slope of the ratio is negative, for $t=t_{\mathrm{turn}}$ the ratio is given by 
$ \mathcal{R}_{k}(t_{\mathrm{turn}})= \frac{3\sqrt{3}}{2} \, \frac{\gamma_{\mathrm{qp}}}{H} \, k^{2}$ , and for $t>t_{\mathrm{turn}}$ the ratio is positive. Therefore the eikonal ratio has a minimum at $t_{\mathrm{turn}}$, with the maximal particle production around this point. After this point the ratio starts to increase again, and we shall soon see that the particle production process will slow down again. 

%
%_figure__figure__figure__figure__figure__figure__figure__figure__figure__figure__figure__figure_
\begin{figure*}[!htb]
\begin{center}
\mbox{
\subfigure[$\,$ $N_{k}(t)$. 
\label{Fig:QP}]{\includegraphics[width=2.8in]{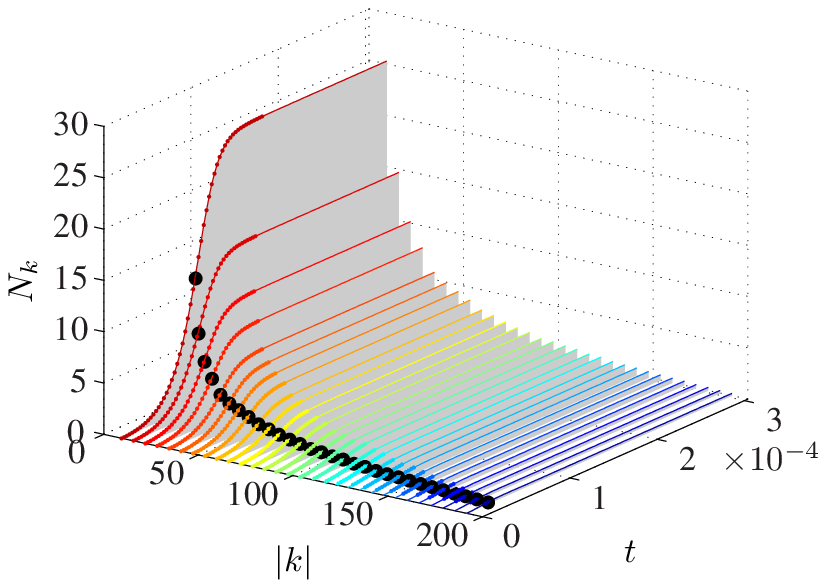}}
\hspace{3mm}
\subfigure[$\,$ $\mathcal R_{k}(t)$. 
\label{Fig:FR}]{\includegraphics[width=2.8in]{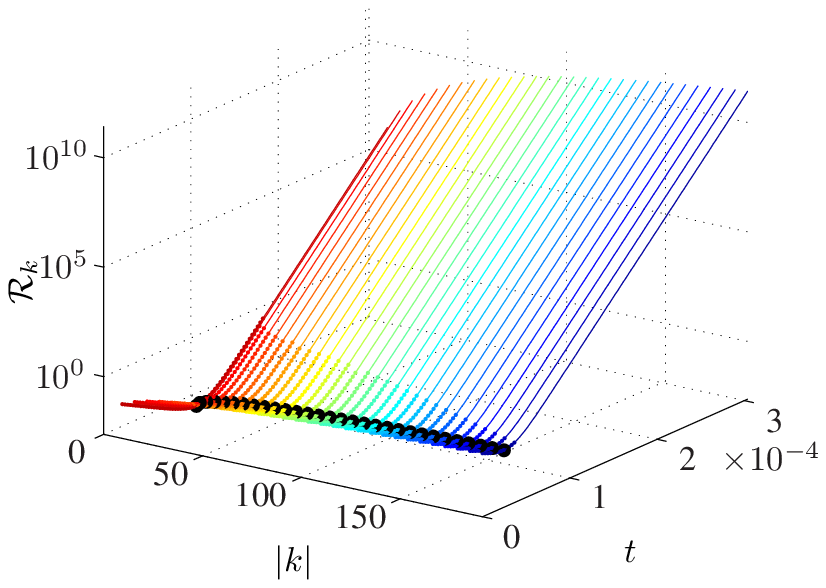}}     
}
\vspace{3mm}
\mbox{
\subfigure[$\,$  $N_{k}(t)$ projected onto the $t$-$N_{k}$ plane.
\label{Fig:FIG_desitter_ts1e-05_4000000_200000_QP_0_0}]{\includegraphics[width=2.8in]{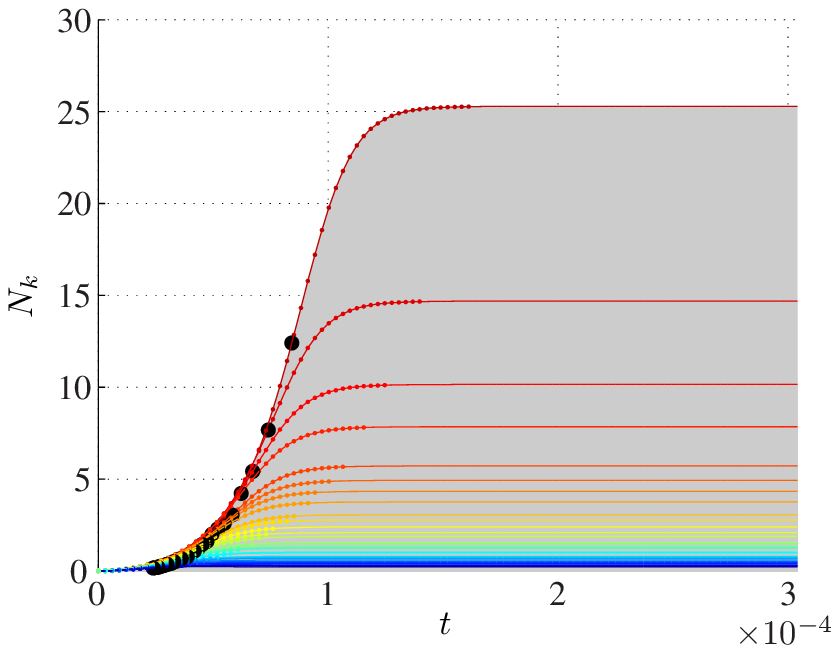}}
\hspace{3mm}
\subfigure[$\,$  $\mathcal R_{k}(t)$ projected onto the $t$-$\mathcal{R}_{k}$ plane.
\label{Fig:FIG_desitter_ts1e-05_4000000_200000_FR_0_0}]{\includegraphics[width=2.8in]{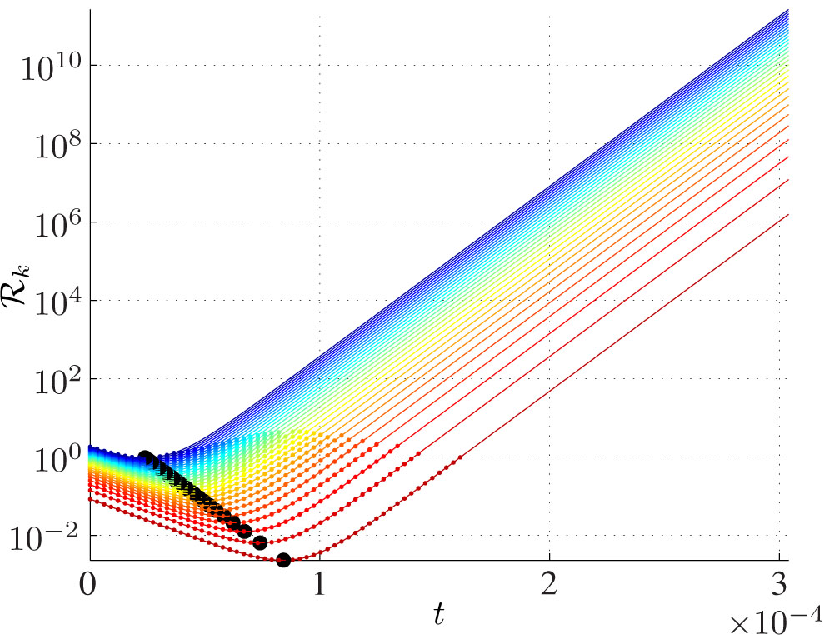}}     
}
\caption[Relationship between quasiparticle production and frequency ration for quantum modes.]{(Colors online only.) In this figure we compare the quasiparticle production per quantum mode (left column) with its frequency ratio $\mathcal{R}_{k}(t)$ (right column), for $t_{s}=1\times10^{-5}$. Parameters are $C_{NL}(\bar{t}=0)=2 \times 10^{5}$, $N_{0}=10^{7}$ and $X=4 \times 10^{6}$. (See~\cite{Jain:2006ki,Weinfurtner:2007ab} for details of the simulations.) The bold plotted dots on the left hand side indicate that the frequency ratio is below one, hence the quantum mode corresponds to a super-Hubble horizon mode. On the right hand side we have indicated with the bold dots when a change in the mode occupation number is above a certain threshold --- here $\Delta N_k \geqslant 0.004$ --- roughly to filter out quantum noise fluctuations. }
\label{Fig:desitter_ts1e-05_4000000_200000_QP_FR}
\end{center}
\end{figure*}
%_figure__figure__figure__figure__figure__figure__figure__figure__figure__figure__figure__figure_
%

To qualitatively describe the particle production process in our specific rainbow spacetime, we suggest the following terminology: 
\begin{description}
\item[$t\to - \infty$:] {At early times almost all modes are ``sub-Hubble-horizon'' modes, and the particle production process is negligible. The modes oscillate with much higher frequencies than their corresponding Hubble frequencies, that is $\mathcal{R}_{k}(t) \gg 1$.}
\item[$t \sim t_{ \mathrm{turn} } $:] {As times goes on the mode frequencies are decreasing, while at the same time the rainbow Hubble frequencies are decreasing as well. Nevertheless, the ratio between them exhibits a minimum at $t_{\mathrm{turn}}$, where the particle production process is expected to be maximal. 
Even if the particle production process is maximal, this does not necessarily imply that the quantity of particle production is noticeable; the modes also need to be ``super-Hubble-horizon'' modes, or in more accurate terminology, we require $\mathcal{R}_{k}(t_{\mathrm{turn}})\ll 1$.}
\item[$t \sim t_{ \mathrm{crossing} } $:] {If there exists a time $t=t_{\mathrm{crossing}}$, such that $R_{k}(t_{\mathrm{crossing}})\sim 1$, where a mode $k$ crosses the ``Hubble horizon'', then there will be a \emph{second} time $t=t_{\mathrm{re-entering}}$, where the mode $k$ re-enters the ``Hubble horizon'', and  $R_{k}(t_{\mathrm{re-entering}})\sim 1$. We suggest that it is useful to adopt the following terminology to describe the behavior of the modes: ``freezing of the mode $k$'' occurs in the time period $t_{\mathrm{crossing}} < t < t_{\mathrm{turn}}$, whereas ``melting of the mode $k$'' occurs during $t_{\mathrm{turn}} < t < t_{\mathrm{re-entering}}$.}
\end{description}

Another novelty in our qualitative understanding of the particle production process in our FRW rainbow-spacetime, is the connection with the condensed matter point of view:
The minimum of the ratio $\mathcal{R}_{k}(t_{\mathrm{turn}})$ for a mode $k$ occurs at
\begin{equation}
\exp(-2Ht) - 2(k/K)^{2}=0 \quad \rightarrow \quad k=\frac{1}{\sqrt{2} \, \ell_{\mathrm{Planck}}(t)}  \, . 
\end{equation}
This quantity also appears in the context of conventional condensed matter physics, where it is defined as the crossover between the phonon and free-particle region. This borderline, the inverse of the healing, or coherence length $\xi$~\cite{Pethick:2001aa}, is given by
\begin{equation}
\xi^{-2} = \frac{2\, m \,n_{0} \, U(t)}{\hbar^{2}} = \frac{1}{2} \, \frac{4 m^{2}}{\hbar^{2}} \, \frac{n_{0} \, U(t)}{m} =\frac{1} {2 \,  \ell_{\mathrm{Planck}}(t)^{2} } \, ,
\end{equation}
which indicates where each mode starts to decouple from the spacetime. That is, each mode can experience particle production, until it becomes free-particle like, and this behaviour starts to kick on near the Planck scale.
Hence, in terms of the microscopic physics of a BEC we have a natural understanding of $\mathcal{R}_{k}(t_{\mathrm{turn}})$. For a detailed treatment of the numerical simulations please see~\cite{Jain:2006ki}.

To show the quantitative correlation between the modified frequency ratio, $\mathcal{R}_k(t)$, with particle production in our specific rainbow de Sitter spacetime, we have plotted the ratio for several $k$-modes as a function of time, and compared them to number occupation plots, see Fig.~\ref{Fig:desitter_ts1e-05_4000000_200000_QP_FR}.
This figure compares the change in the mode occupation number in each mode on the left side, with frequency ratio $\mathcal{R}_{k}(t)$ of the mode on the right side,  for the scaling time $t_{s}=\times 10^{-5}$.
Each coloured line on the left hand side indicates the occupation number in the mode $k$ as a function of time. On the right hand side we have plotted the frequency ratio $\mathcal{R}_{k}(t)$ for each of those modes with a different colour (online only); gradually changing from infrared modes (dark red), to ultraviolet modes (dark blue).

The black dots in the figures to the left indicate when the modes cross over from phononic to trans-phononic behavior, i.e., where each mode starts to decouple from the emergent spacetime. We can see that the black dots are located where the frequency ratio has its minimum; see Figs.~\ref{Fig:desitter_ts1e-05_4000000_200000_QP_FR}.\\

It is also noteworthy that the commutation relation for the perturbations, in terms of the emergent scalar field and its conjugate momentum with respect to the preferred rest-frame --- the laboratory frame --- exhibits an explicit momentum-dependence, $ \left[ \partial_{t} \hat \theta_{k}, \hat \theta_{k} \right] \sim U  + k^{2}$. (Here $\theta_{k}$ is the phase fluctuation operator in momentum-space.) Consequently, the size of the quantum fluctuation is growing with the momentum of the quantum field mode. (In a sense, Planck's constant is being generalized in a momentum-dependent manner.) That is, the precision with which you can measure the macroscopic field variable $\langle  \hat \theta_{k} \rangle$ decreases as the momentum increases. 
While the emergent spacetime picture is necessary to understand the time-dependent commutation relations --- in terms of the field operator and its conjugate momentum on a time-dependent classical background --- the fact that particle production process tends to zero during inflation is naturally explained from a condensed matter physics point of view, as it is related to the crossover between phononic and trans-phononic quasi-particle excitations.

%%%%%%%%%%%%%%%%%%%%%%%%%%%%%%%%%%%%%%%%%%%%%%%%%%%
%
\section{Conclusions\label{Sec:Conclusions}}
%
%%%%%%%%%%%%%%%%%%%%%%%%%%%%%%%%%%%%%%%%%%%%%%%%%%%
What have we learnt from the model presented above? The short and sincere answer is, that we have studied in depth a particular molecular system that shares some features with quantum field theory in curved-spacetime. The viable regime, where the analogy is sufficiently good to mimic the behavior of quantum modes during an inflationary epoch, is rather narrow. While we are able to connect a certain parameter regime of the Bose--Einstein condensate model with semi-classical quantum gravity there is no obvious evidence for its significance for ``full'' quantum gravity. Any sensible theory of quantum gravity has to accommodate dynamics for the gravitational field which, in the appropriate limit, should reduce to those of general relativity (or at least be sufficiently close to the latter, see modified theories of gravity such as~\cite{Bekenstein:2004ne, Jacobson:2008le, Lue:2004fe, Maartens:2004hb, Sotiriou:2007uk}). Therefore the analogue model investigated above is not a suitable candidate for ``full'' quantum gravity. Thankfully, however, the breakdown of the analogy is to a very good extent uncorrelated with the applicability of the linearization scheme, and (for questions of cosmological particle production that we are primarily interested in) we are not picking-up deviations caused by back-reaction effects. \\

For the sake of the argument let us go further by concentrating on the relevance of our model for the phenomenological side of quantum gravity. Quantum gravity phenomenology is the attempt to reduce quantum gravity to a series of high-precision effects manifesting themselves in minor extensions of our currently known theories. (It is nothing else then, than an existentialist\footnote{Jean-Paul Sartre, Being and nothingness.} approach towards quantum gravity.) One assumes that quantum gravity, as the fundamental theory behind quantum field theory and general relativity, cannot hide completely, and will eventually reveal its existence in terms of modifications to one or both of these theories. Lacking a theory of quantum gravity, we are left with \textit{physically reasonable} guesses about the phenomenological side of quantum gravity.
In this spirit analogue models for gravity/ emergent spacetimes might be thought of as an inspiration for quantum gravity phenomenology. They are real-life examples for the idea, that independent of the actual system (e.g. electromagnetic waveguide, superfluid helium, condensed Bosons, water)  a notion of spacetime exists in the infrared limit, hiding its actual fundamental nature. What is more, the first corrections accounting for the specific substructure enter the quantum field equations in a familiar manner, as expected in many effective field theories. As to the sense or nonsense of emergent rainbow spacetimes, there is no physics restriction in the present model that would prevent us from  considering  the ``Planck-modifications'' as part of the geometry. \\

In conclusion,  we can only hope that the "deja vu effect" in physics is to some extent applicable for quantum gravity, and to let nature decide what is physically reasonable.  

\section*{Acknowledgments}

The research reported in this work was supported by the Marsden Fund administered by the Royal Society of New Zealand (MV \& SW, contract VUW-0606). Additional support was provided by two Victoria University PhD completion scholarships (SW \& PJ).  Additionally SW and MV would like to particularly thank the conference organizers for their hospitality. CWG was supported by the New Zealand Foundation for Research Science and Technology under the contract NERF-UOOX0703 and by the Marsden Fund under contract UOO-0590. Additionally, SW and MV would like to thank the conference organizers. In particular SW wishes to thank Stefano Liberati and Paola Roncaglia for their hospitality, and Thomas Sotiriou and Bill Unruh for their thoughtful comments and suggestions. 

\clearpage
\bibliographystyle{habbrv}

\end{document}